\documentclass[twocolumn,aps,prl,showpacs,floatfix]{revtex4} % ketoszlopos
\usepackage{graphics}
\usepackage{epsfig}
\usepackage{indentfirst}
\usepackage{hhline}

\begin{document}
%%%%%
\title{Dynamic scaling of fronts in the quantum XX chain}

\author{V. Hunyadi$^{1}$, Z. R\'acz$^{2}$, and L. Sasv\'ari$^{1}$}
\affiliation{${}^1$Department of Physics of Complex Systems, E\"otv\"os University,
P.O.Box 32, H-1518 Budapest, Hungary\\
${}^2$Institute for Theoretical Physics - HAS, E\"otv\"os University,
1117 Budapest, P\'azm\'any s\'et\'any 1/a, Hungary}

\date{\today}

\begin{abstract}
The dynamics of the transverse magnetization in the
zero-temperature $XX$ chain is studied with emphasis on
fronts emerging from steplike initial magnetization
profiles. The fronts move with fixed velocity and display
a staircase like internal structure whose dynamic scaling is explored
both analytically and numerically. The front region is found
to spread with time sub-diffusively with
the height and the width of the staircase
steps scaling as $t^{-1/3}$ and $t^{1/3}$, respectively.
The areas under the steps are independent of time, thus the
magnetization relaxes in quantized "steps" of spin-flips.
\end{abstract}
\maketitle
%%%%%
%\section{Introduction}
%%%%%%%%%%%%%%%%%%%%%%%%%%
Fronts often emerge in relaxation processes.
Simple examples are the domain walls generated in phase-separation
dynamics \cite{Bray-review} but there is a long list of fronts
(also called shocks, active zones, reaction zones, etc.)
resulting from unstable dynamics in various physical,
chemical and biological systems \cite{{Cross-Hoh},{Murray}}.
Furthermore, fronts can also emerge due to
the initial spatial separation of stable and unstable
states \cite{Saarloos}.

Apart from being the instruments of relaxation, the importance of
fronts is also due to their "catalytic" nature. Namely,
structures are built in them \cite{Plis-R-84} and patterns often emerge
in the wake of moving fronts \cite{Galfi}. Thus it is not surprising that
much effort has been devoted to the description of their
spatio-temporal structure \cite{{Cross-Hoh},{Saarloos}}.

In contrast to fronts in classical systems, not much
is known about quantum fronts. The examples
we are aware of are restricted to quantum spin chains:
fronts or shocks have been seen in the
$XX$ model \cite{{xx-aram},{xx-front},{Ogata2}}, in the transverse
Ising chain \cite{{Igloi},{karevski}}, as well as
in the Heisenberg model \cite{{salerno},{popkov}}.
The detailed structures of the
front regions, however, have not been elucidated even in these cases.

Having in mind the importance of the front dynamics,
we set out to investigate the fronts emerging in
the quantum $XX$ chain. The transverse magnetization in this model is
conserved and its relaxation is known to be
governed by fronts provided an initial
state with a spatial separation of distinct magnetizations ($\pm m_0$)
is prepared \cite{xx-front}.
The problem with such initial condition is exactly solvable
and thus we can follow the
evolution of the spatio-temporal structure of the front
in detail.

The main result of our calculations is that, in addition
to the known \cite{xx-front} global scaling
(finite front velocity and well defined magnetization profile in the
$t\to \infty$ limit), there exists a dynamic
scaling regime in the front, and the associated scaling function exposes
a staircase structure in the magnetization profile. An important
feature of the steps in the staircase is that while their height
decreases as $t^{-1/3}$ and their width increases as $t^{1/3}$, the
areas under the steps are constants and, furthermore, the
constants are the same for all the steps.
The value of the constant (twice the magnetic moment of a spin)
indicates that a step carries a reversed spin with respect to the
aligned initial state. These reversed spins move with the front velocity,
keep their identity with respect to other reversed spins, and their
spatial spread is sub-diffusive. Thus the magnetization relaxes
in well defined quantized (in space and in time) "steps".

Before turning to the calculations, we note that our results suggest
that the quantum fronts and, in particular, the steps carrying
a unit of spin-flips, may be envisioned as ingredients in controlled
transport of bitwise information in magnetic nanostructures.

\par The system we investigate is the XX chain defined
by the Hamiltonian
\begin{equation}
{\hat H}=-\sum_{j=-N+1}^{N-1}\left(S_j^xS_{j+1}^x + S_j^yS_{j+1}^y\right)
\end{equation}
where the spins $S_j^\alpha$ $(\alpha=x,y,z)$ are $1/2$ times the
Pauli matrices situated at the sites $[j=0,\pm 1,...,\pm (N-1),N]$
of a one-dimensional lattice. Free boundary conditions are used,
and the thermodynamic limit $N\to \infty$ is assumed throughout the paper.
The physical quantities are measured in their
natural units: energy in $J\hbar^2$ where J is
the nearest-neighbor coupling, magnetization in $\hbar$,
length in units of the lattice constant $a$,
and time in $1/J\hbar$.

\begin{figure}[htb]
\includegraphics[width=8cm]{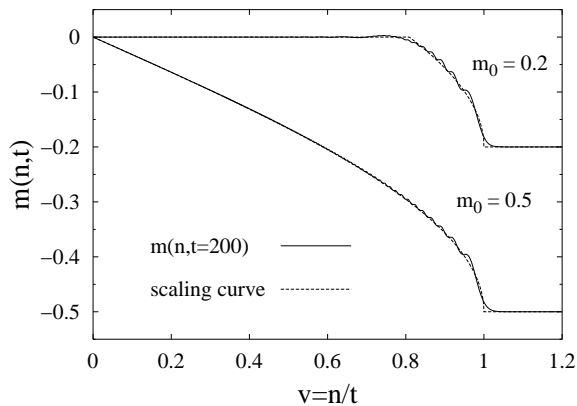}
\caption{Magnetization profiles emerging from steplike initial conditions
[eqs. (\ref{fimax}) and (\ref{alt})]. The large-time
scaling limit is shown by dashed lines  [eq.~(\ref{skmagn}) for $m_0=0.5$].}
\label{a_skmagn}
\end{figure}

Our principal aim is to investigate the time evolution of the
(globally conserved) transverse magnetization
\begin{equation}
m(n,t)=\langle\varphi |\,S_n^z(t)\,|\varphi\rangle
\end{equation}
emerging from an initial state
$|\,\varphi\rangle$ with a steplike magnetization profile
\begin{equation}
\mid\varphi\,\rangle \, =\hspace{3pt} \mid \hspace{-8pt}\stackrel{-N+1}{\uparrow}\dots
\hspace{5pt}{\uparrow}
\hspace{-3pt}\stackrel{0}{\uparrow} \stackrel{1}{\downarrow} \downarrow\hspace{5pt}\dots
\hspace{5pt}\stackrel{N}{\downarrow}\hspace{1pt} \rangle \, . \label{fimax}
\end{equation}
Since the dynamics following from ${\hat H}$ can be described \cite{jordan,lieb}
in terms of local fermionic operators $(c_n, c_n^\dagger)$ whose Fourier transforms
diagonalize ${\hat H}$, and since $S_n^z$ can be expressed
through the local fermionic operators as $S_n^z=c_n^\dagger c_n-1/2$,
the evaluation of $\langle\varphi|\,S_n^z(t)\,|\varphi\rangle$ is a relatively simple exercise.
The calculations have been carried out in \cite{xx-front} with the result
for $n\ge 1$ given through the Bessel functions of the first kind \cite{Abr-Steg}
\begin{equation}
m(n,t) = -\frac12 J_0^2(t)- \sum_{l=1}^{n-1}J_l^2(t)
\label{magnveges}
\end{equation}
while the expression for $n\le 0$ is obtained
from symmetry considerations $m(n,t)=-m(-n+1,t)$.
The global scaling of $m(n,t)$ emerges in the
$n\rightarrow +\infty$, $t\rightarrow+\infty$ and $n/t=\mbox{finite}$
limit where $m(n,t)$ can be written in a scaling form
$m(n,t)\to\Phi\left(\frac{n}{t}\right)$, and the scaling function
$\Phi\left(v\right)$ is given \cite{xx-front} by
\begin{equation}
\Phi(v)=
\left\{
\begin{array}{ll}
\displaystyle -{\pi}^{-1}\arcsin(v) & \mbox{for } 0\le v\le 1,\\
\\
\displaystyle -1/2 & \mbox{for } v\ge 1.
\end{array}
\right.
\label{skmagn}
\end{equation}
The $m_0=0.5$ curve in Fig.\ref{a_skmagn} shows $\Phi(v)$ together with the
shape of $m(n,t)$ plotted
against $v=n/t$ at finite time $(t=200)$. As one can see, the magnetization
displays a staircase structure near the edge of the front $(n/t\approx 1)$. Our
main concern will be the scaling properties (both in space and time)
of this staircase structure.
%%%%%%%%%%%%

\begin{figure}[h!]
\centerline{
\resizebox{85mm}{!}{
\includegraphics{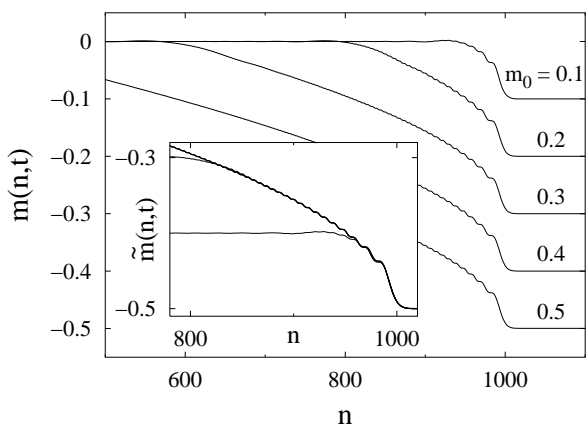}
} %resizebox
}
\caption{Magnetization profiles at $t=10^3$ for different initial values $m_0$
[eq.(\ref{alt})]. The inset with the curves shifted vertically down
($\tilde m =m-0.5+m_0$) demonstrates that the steps in the front region
are independent of the value of $m_0$.}
\label{a_altalanos}
\end{figure}

%%%%%%%%%%%%
For possible applications, it may be important that the front properties
were tunable. Possibility for tuning can be seen by solving the problem
for more general steplike initial states
\begin{equation}
\langle \,\varphi_{m_0}\mid S_n^z(0)\mid \varphi_{m_0}\, \rangle =
\left
\{\begin{array}{ll}
m_0 & \mbox{ for } -N < n \le 0\\
-m_0 & \mbox{ for }  \,\, 1\le n \le N \,\, .
\end{array}
\right.
\label{alt}
\end{equation}
where $\mid \varphi_{m_0}\,\rangle$ is constructed by joining two half chains which
are the ground states of the $XX$ model at magnetizations $\pm m_0$.
The expression to be analyzed for this case can be taken from Ref.\cite{xx-front},
eq.(12).
It is more involved but its numerical evaluation does not pose difficulties.
We shall not describe the numerical work but results for $m_0\not= 0$ will be
discussed and displayed (Figs.\ref{a_skmagn} and \ref{a_altalanos}).

\begin{figure}[htb]
\includegraphics[width=8cm]{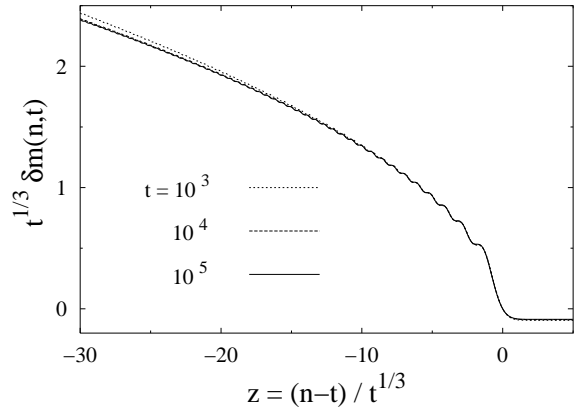}
\caption{Scaling in the front region. The magnetization measured form
$m(t,t)$ is magnified as $t^{1/3}\cdot\delta m(n,t)$ while the
distance from the front position is scaled as $z=(n-t)/t^{1/3}$.
The analytically derived scaling limit
[eq.(\ref{scalingfunc})] is indistinguishable from
the $t=10^5$ curve in the $z$-range plotted.}
\label{a_magnai}
\end{figure}

The present work on the dynamic scaling of the magnetization
profile around the edge of the front $(n\approx t)$ originates
from our numerical studies of the
deviation of the magnetization $\delta m(n,t)=m(n,t)-m(t,t)$ from its
front value $m(t,t)$. As one can observe
on Fig.\ref{a_magnai}, a well defined scaling function emerges
in the limit $t\to \infty$ provided the
magnetization and the region around $n\approx t$ are scaled by
$t^{1/3}$ and $t^{-1/3}$, respectively.
%%%%%
%%%%%

An analytic understanding of the scaling seen in Fig.\ref{a_magnai} can be
developed by first using (\ref{magnveges}) to write $\delta m(n,t)$ as
\begin{eqnarray}
\delta m(n,t)=\left\{
\begin{array}{lr}
\displaystyle +\sum\limits_{l=n}^{t-1} J_l^2(t) & \mbox{ for }n<t, \\
\displaystyle\quad \quad\quad 0 & \mbox{ for } n=t, \\
\displaystyle -\sum\limits_{l=t}^{n-1} J_l^2(t) & \mbox{ for } n>t.
\end{array}\right.
\label{magnkul}
\end{eqnarray}
Next, one can observe that, in what we call the scaling regime
($t\to \infty$, $n\to \infty$, and $|t-n|\approx t^{1/3}$),
the sums above contain Bessel functions whose index ($l$) and argument ($t$)
differ from each other at most by $|l-t|\sim {\cal O}(t^{1/3})$. This
suggests the use of the following asymptotic
expansion of Bessel function \cite{Abr-Steg}:
\begin{equation}
J_n(n+zn^{1/3})\approx 2^{1/3}n^{-1/3}{\rm Ai}(-2^{1/3}z) + {\cal O}
(n^{-1})
\label{airy}
\end{equation}
where ${\rm Ai}(z)$ is the Airy function. Indeed, in the scaling regime,
the terms in the sums in eq.(\ref{magnkul}) can be written as
\begin{equation}
J_l^2(t)\approx \frac{2^{2/3}}{l^{2/3}}
{\rm Ai}^2\left(2^{1/3}\frac{l-t}{l^{1/3}}\right)\approx \frac{1}{t^{2/3}}
F\left(\frac{l-t}{t^{1/3}} \right)
\label{ketharmad}
\end{equation}
where $F(y)=2^{2/3}{\rm Ai}^2(2^{1/3}y)$. Note that
the last expression in (\ref{ketharmad}) is obtained by replacing
$l$ by $t$ in the factors $l^{1/3}$ and $l^{2/3}$, an approximation
that is valid in the scaling regime.

\par The last step in deriving $\delta m(n,t)$ is the calculation of
the sums in (\ref{magnkul}) using eq.(\ref{ketharmad}).
In the scaling regime, the sums can be replaced by integrals and,
introducing the variable $y=(l-t)/t^{1/3}$, both sums yield the same
expression. As a consequence, we obtain a single scaling form for
$n>t$ as well as for $n\le t$
\begin{equation}
\delta m(n,t) =
\frac{1}{t^{1/3}} G\left(\frac{n-t}{t^{1/3}}\right).
\label{skalaz}
\end{equation}
where the scaling function $G(z)$ is explicitely
given by the following integral
\begin{equation}
G(z) =-\int\limits_{0}^{z} \mathrm dyF(y)=
-2^{2/3}\int\limits_{0}^{z}\mathrm dy{\rm Ai}^2(2^{1/3}y)\, .
\label{scalingfunc}
\end{equation}
The scaling form (\ref{skalaz}) and the scaling function (\ref{scalingfunc})
are our central result. The staircase shape of the front follows from
(\ref{scalingfunc}) while the scaling form (\ref{skalaz}) shows that
the width of the steps increases as $w\sim t^{1/3}$
while their heights (measured from the $z\to\infty$ level)
decrease as $h\sim t^{-1/3}$.

An important feature of the above scaling is that since $w\times h\sim t^0$,
the areas under the steps are constants. Thus a step
carries a given amount of magnetic moment $\mu_s=w_s\times h_s$ where
subscript $s$ denotes the $s^{\rm th}$ step counted from the edge of the front.
In principle, $\mu_s$
could depend on $s$ but their numerical evaluation
for the steps near the edge of the front $(1\le s \le 8)$
(see Table \ref{t_ter1})
strongly suggests that $\mu_s=1$ (the deviations from $1$
can be attributed to the finiteness of the sums at finite $t$).

%%%%%
\begin{table}[h!]
\begin{center}
\begin{tabular}{|c|c|c|c|}
\hline
step $\rm n^o$:$s$ &$\mu_s$(t=$10^3$)&$\mu_s$(t=$10^4$)&$\mu_s$(t=$10^5$)\\
\hline
1 & 1.0183 & 1.0420&1.0379 \\
2 & 0.9707 & 1.0154&1.0094 \\
3 & 1.0371 & 1.0080&0.9934 \\
4 & 0.9418 & 0.9710&1.0124 \\
5 & 0.9022 & 0.9970&0.9986 \\
6 & 0.9588 & 1.0070&0.9860 \\
7 & 1.0123 & 0.9434&1.0128 \\
8 & 0.9286 & 0.9866&1.0037 \\
\hline
\end{tabular}
\end{center}
\caption{Magnetic moments carried by the steps
(area below the steps in the magnetization profile) at finite
times ($t=10^3, \,10^4, \, 10^5$).
The borders of the steps were defined by the inflection points
in the magnetization curve.}
\label{t_ter1}
\end{table}
The $\mu_s=1$ result can be derived for large-order ($s\gg 1$) steps.
Indeed, as can be seen from Fig.\ref{a_skmagn}, the height of the step can be
estimated from the global scaling function $m(n,t)=-\pi^{-1}\arcsin{(1-z/t^{2/3})}
\approx -1/2+\pi^{-1}\sqrt{2|z|}t^{-1/3}$, and thus
$\delta m= h_s=\pi^{-1}\sqrt{2|z_s|}t^{-1/3}$
where $z_s$ is the scaling variable at the $s^{\rm th}$ step. Its value
is found by finding the $s^{\rm th}$ zero of $G^\prime(z)$
which, in turn [see eq.(\ref{scalingfunc})], is obtained from the $s^{\rm th}$
solution of the ${\rm Ai}(2^{1/3}z)=0$ equation.
The large $s$ asymptotic of the $s^{\rm th}$ zero is given \cite{Abr-Steg}
by $|z_s|\approx (3\pi s)^{2/3}/2$ and thus
\begin{equation}
h_s=[3 s/(\pi^2t)]^{1/3} \, .
\label{hs}
\end{equation}
We need to find now the width of the $s^{\rm th}$ step. Defining the
borders of the steps as the consecutive inflection points on the staircase,
one can see from (\ref{scalingfunc}) that the width of the
step is given by the consecutive zeros of ${\rm Ai^\prime}(2^{1/3}\tilde z_s)=0$.
The large $s$ asymptotic of these zeros are known again \cite{Abr-Steg} and we
find $|\tilde z_{s+1}-\tilde z_{s} |\approx [\pi^2/(3s)]^{1/3}$. We should remember
now that $h_s$ was calculated for the original (unscaled) magnetization, and the
width of the step must also be obtained in unscaled spatial coordinates.
Consequently, we should scale $|\tilde z_{s+1}-\tilde z_{s} |$ by $t^{1/3}$ and
thus
\begin{equation}
w_s=t^{1/3}|\tilde z_{s+1}-\tilde z_{s} |\approx[\pi^2t/(3s)]^{1/3} \, .
\label{ws}
\end{equation}
Comparing eqs.(\ref{hs}) and (\ref{ws}) we see that $h_s\cdot w_s=1$ thus
arriving at an important property of the staircase structure, namely, the
steps of the staircase carry a unit of magnetization.

In order to see the relevance of the above result and to develop a picture
about it, let us consider the magnetization flux in the front region.
The total transverse magnetization $M^z=\sum_ns_n^z$ is conserved and so
one can define the local magnetization flux,
$
{\hat \mathrm {J}}_n^z=S_{n}^yS_{n+1}^x - S_n^xS_{n+1}^y,
$
which is related to the local magnetization through the continuity equation.
Thus, not surprisingly, one can derive a simple expression for the time evolution
of the expectation value of ${\hat \mathrm {J}}_n^z$ as well
\begin{equation}
j(n,t)\equiv\langle \, \varphi \, | \, {\hat \mathrm {J}}_n^z(t)\, | \, \varphi \, \rangle
= \sum_{l=n}^{\infty} J_l(t) J_{l+1}(t) .
\label{aramveges}
\end{equation}
The analysis of $\delta j(n,t)=j(n,t)-j(t,t)$ parallels that of
$\delta m(n,t)$ and yields the same scaling structure in the front
\begin{equation}
\delta j(n,t)=\frac{1}{t^{1/3}} G\left(\frac{n-t}{t^{1/3}}\right)=\delta m(n,t).
\end{equation}
Adding $G(\infty)/t^{1/3}$ to both sides of the above equations, and
taking into account that $j(n\to \infty,t)=0$ and $m(n\to \infty,t)=-1/2$ and,
furthermore, remembering that the front moves with velocity $v=1$, we can write
the above equations in a form that is easy to interpret
\begin{equation}
j(n,t)=v\left[ m(n,t)-(-\frac{1}{2})\right ] \, .
\end{equation}
Since the sum of $[m(n,t)-(-1/2)]$ for an interval gives the number of up-spins
in that interval, the meaning of the unit area under under the
steps is as follows. The steps represent spatial intervals in which
a single up-spin is spread in the sea of down-spins. These up-spins behave like
particles, they move with velocity $v=1$ and provide
the magnetization flux. Furthermore, they are localized in the sense
that their spatial spread
is sub-diffusive ($\sim t^{1/3}$) instead of the usual quantum mechanical
spreading ($\sim t^{1/2}$). The picture thus emerging is somewhat
reminiscent of the
hard-core bosons description of the $XX$ model near its
critical external field \cite{Sachdev}.

\par From the point of view of possible applications, it is
important that the staircase structure can be observed for other initial
conditions as well. Using initial states $| \varphi_{m_0} \rangle$
with $\pm m_0$ steplike profile as described in eq.(\ref{alt}),
we found numerically that the staircase emerges again
(see Fig.\ref{a_altalanos}). Furthermore, the steps of the staircase
were found to have the same size
independently of values of $m_0$ (see inset in Fig.~\ref{a_altalanos}).
What is varied with $m_0$ is the number of steps ${N}(t, m_0)$
in the staircase.
%%%%%
%%%%%

We can estimate ${N}(t, m_0)$ by
assuming (on the basis of numerical solutions) that scaling extends
to the whole front region, up to the point
where $m$ reaches the steady state value of $m=0$.
Then $N(t, m_0)$ is found by locating the $N^{th}$ step which has a
width $w_{N}$ such that
a single spin flip changes the magnetization from $-m_0$ to $0$.
Since $m_0\cdot w_N$ is the difference between the number of up- and
down-spins in the $N^{th}$ step, the above consideration gives the
condition $-m_0 \cdot w_N+1=0$. For large $N$, eq.(\ref{ws})
gives the width $w_N$, and we obtain
\begin{equation}
N(t, m_0)\approx \pi^2m_0^3t/3 \, .
\label{nosteps}
\end{equation}
It should be noted
that eq.(\ref{ws}) is an excellent approximation also for step-numbers
of the order of unity, and so eq.(\ref{nosteps}) gives $N(t,m_0)$ for
small $N$, as well. Indeed, e.g. eq.(\ref{nosteps}) yields $N\approx 3$
for $t=10^3$ and $m_0=0.1$ while the corresponding curve on
Fig.\ref{a_altalanos} displays $N\approx 4$ steps.

An important feature of eq.(\ref{nosteps}) is the strong dependence of
the step number on the initial magnetization $(N\sim m_0^3)$. This gives
a sensitive control over how many steps arrive at a given point at a given
time. Thus, in principle, one can imagine applications with the steps
transferring bits of information in magnetic nanostructures.
The aim of the present study, however, was only to draw attention
to the remarkable features of quantum fronts. Applications would
require answers to a number of non-trivial questions.
First, can fronts with
properties found in the $XX$ model be observed in other spin chains?
In particular, are they present in non-integrable systems? Second,
how does the step structure change
at small but finite temperatures? Finally, are the front properties
robust enough to survive small perturbations arising from impurities?
These are difficult problems but the available analytical and numerical
techniques may be sufficient to tackle them.

In summary, our calculations exposed a dynamic scaling regime
in the motion of fronts in the quantum XX model. The fronts display
a staircase magnetization profile which
has a simple interpretation in terms of single spin-flips spread over
the spatial extent of the steps. Whether these fronts have a direct application
is an open question but their properties are intriguing enough to
look for similar structures in more complex spin-chains.

\acknowledgments

We would like to thank T. Antal, J. Cserti, V. Eisler, A. R\'akos,
G. M. Sch\"utz, and F. van Wijland for useful discussions.
This research has been supported by the Hungarian Academy
of Sciences (Grant No.\ OTKA T029792 and T043734).
%%%%%%%%%%%%%%%%%%%%%%%%%%%%%%%%%%%%%%%%%%%%%%%%%%%%%%%

%%%%%%%%%%%%%%%%%%%%%%%%%%%%%%%%%%%%%%%%%%%%%%%%%%%%%%%
\end{document}